\documentstyle[emulateapj,apjfonts]{article}

\begin{document}
\submitted{To appear in ApJ Letters}
\title{Far-UV Observations of NGC~4151 during the ORFEUS-SPAS~II Mission}
\author{B.\ R.\ Espey\altaffilmark{1}, G.\ A.\ Kriss\altaffilmark{1},
J.\ H.\ Krolik\altaffilmark{1},
W.\ Zheng\altaffilmark{1},
Z.\ Tsvetanov\altaffilmark{1}},
\and
\author{A.\ F.\ Davidsen\altaffilmark{1}}
\affil{Center for Astrophysical Sciences, Department of Physics and Astronomy,
Johns Hopkins University, Baltimore, MD 21218;\\
espey@pha.jhu.edu, gak@pha.jhu.edu,
jhk@pha.jhu.edu,
zheng@pha.jhu.edu,
zlatan@pha.jhu.edu, 
afd@pha.jhu.edu
}

\authoremail{espey@pha.jhu.edu}
\altaffiltext{1}{ORFEUS-SPAS~II Guest Investigator}

\begin{abstract}
We observed the Seyfert 1 galaxy NGC~4151 on eleven occasions at 1--2
day intervals using the Berkeley spectrometer
during the ORFEUS-SPAS~II mission in 1996 November.
The mean spectrum covers 912--1220 \AA\
at $\sim$0.3 \AA\ resolution
with a total exposure of 15,658 seconds.
The mean flux at 1000 \AA\ was
$4.7 \times 10^{-13}~\rm erg~cm^{-2}~s^{-1}~\AA^{-1}$.
We identify the neutral hydrogen absorption with a number of components
that correspond to the velocity distribution of \ion{H}{1} seen in our own
Galaxy as well as features identified in the
\ion{C}{4} $\lambda 1549$ absorption profile by Weymann et al.
The main component of neutral hydrogen in NGC~4151 
has a total column density of $\rm log~N_{HI} = 18.7 \pm 1.5~\rm cm^{-2}$ for a
Doppler parameter $b = 250 \pm 50~\rm km~s^{-1}$, and it covers $84 \pm 6$\%
of the source.
This is consistent with previous results obtained with the
Hopkins Ultraviolet Telescope.
Other intrinsic far-UV absorption features are not resolved, but the
\ion{C}{3}$^*$ $\lambda 1176$ absorption line has a significantly higher
blueshift relative to NGC~4151 than the \ion{C}{3} $\lambda 977$ resonance line.
This implies that the highest velocity region of the outflowing gas
has the highest density.
Variations in the equivalent width of the \ion{C}{3}$^*$ $\lambda 1176$
absorption line anticorrelate with continuum variations on timescales of days.
For an ionization timescale of $<$1 day, we set an upper limit of 25 pc on the
distance of the absorbing gas from the central source.
The \ion{O}{6} $\lambda 1034$ and \ion{He}{2} $\lambda 1085$ emission lines
also vary on timescales of 1--2 days, but their response to the continuum
variations is complex.  For some continuum variations they show no response,
while for others the response is instantaneous to the limit of our sampling
interval.
\end{abstract}


\section{INTRODUCTION}

The Seyfert 1 galaxy NGC~4151 has been a favorite of observers
wishing to use variability as a probe of the inner workings of active
galactic nuclei (AGN).
NGC~4151's ultraviolet variability was first noted in a rocket flight
(\cite{Hartig79}).
Numerous campaigns using the {\it International Ultraviolet Explorer}
(IUE) have since monitored the continuum, emission lines and absorption lines
(\cite{Ulrich84}; \cite{Bromage85}; \cite{Clavel92};
\cite{Crenshaw96}; \cite{Edelson96}).
IUE studies of the absorption lines (\cite{Bromage85}) noted a tendency
for the equivalent widths of the high ionization lines (e.g., \ion{C}{4})
to correlate directly with variations in the continuum while
low ionization lines such as \ion{C}{2} were anticorrelated.
In contrast,
monitoring of the \ion{C}{4} absorption line a decade later using the
Goddard High Resolution Spectrograph (GHRS)
on the Hubble Space Telescope (HST)
at moderate spectral resolution (R$\sim$15,000)
showed no variation in strength or shape
over a baseline of several years (\cite{Weymann97}).

Significant UV continuum and emission-line variability on timescales
as short as 1--2 days were seen in the continuous monitoring of NGC~4151 with
IUE in 1994 December (\cite{Crenshaw96}).
Monitoring with the Hopkins Ultraviolet Telescope (HUT) during the Astro-2
mission over the 912--1820 \AA\ band at $\sim 2$-day intervals
showed no change in the \ion{C}{4} equivalent width, but revealed substantial
variations in the \ion{H}{1} column and in low-ionization ions such as
\ion{C}{3} and \ion{Si}{4} that were
correlated with variations in the continuum (\cite{Kriss96}; \cite{Kriss98}).
With the low resolution (R$\sim$300) HUT data, however, we could not
identify which of the several components in the complex \ion{C}{4} absorption
profile was associated with these variations.

The HUT Astro-1 and Astro-2 observations of NGC~4151 revealed numerous
absorption lines over a wide range of ionization states in the 912--1200 \AA\ 
band (\cite{Kriss92}; \cite{Kriss95}).
Our observations with the Berkeley spectrometer on ORFEUS-SPAS~II
were intended to make use of its higher resolution (R$\sim$3300) relative to HUT
to study the structure and time response of the absorption line systems
in greater detail.
We present here the mean spectrum and key features of the variability
we saw during the ORFEUS-SPAS~II mission.
Full details of the individual observations will be described in a subsequent
publication.

\section{OBSERVATIONS AND DATA REDUCTION}

We obtained spectra of NGC~4151 at eleven epochs with the Berkeley spectrograph
during the ORFEUS-SPAS~II mission in 1996 November/December.
Exposure times ranged from 600 to 2200 s, with a total exposure of 15,658 s.
The general design of the Berkeley spectrograph is discussed by
Hurwitz \& Bowyer (1986\markcite{Hurwitz86}, 1996\markcite{Hurwitz96}),
while calibration and performance for the ORFEUS-SPAS~II mission are described
by Hurwitz et al.\ (1997).\markcite{Hurwitz97a}

The Berkeley spectrometer has a 2-D detector, so the NGC~4151 data
were obtained through a 26\arcsec\ diameter aperture simultaneously with
airglow spectra through a larger aperture offset by 2.4\arcmin.
The signal-to-noise ratio varies greatly with wavelength, and it is highest
near the \ion{O}{6} and Ly$\alpha$\ emission lines.
Due to a reflection in the Berkeley spectrometer, undispersed Ly$\alpha$
airglow produces a rapid increase in the background level below $\sim$950 \AA.
Correction for this and other instrumental features are discussed in
Hurwitz et al.\ (1997)\markcite{Hurwitz97a}, and our data were extracted and
calibrated using the prescription
recommended by the Berkeley spectrometer team. 

To maximize the signal-to-noise ratio (S/N), we binned these basic spectra into
larger wavelength bins of 0.15 \AA,
roughly half a resolution element (\cite{Hurwitz97a}).
To correct the wavelength scale for slight inaccuracies in telescope pointing,
we shifted the object spectra to place the Galactic \ion{Si}{2}
$\lambda 1190$ and $\lambda 1193$ absorption features
at a heliocentric velocity of $60~\rm km~s^{-1}$
(the Galactic \ion{H}{1} velocity determined by \cite{Murphy97}).
From a comparison among all the observed Galactic absorption features in the
spectrum, we estimate that our wavelength scale is accurate
to $40~\rm km~s^{-1}$ (1$\sigma$).

The contemporaneous airglow spectra were extracted
in exactly the same manner as the target data.
Airglow emission lines were fitted using Gaussian components.
Comparison of the line widths of
isolated emission lines through both the airglow aperture and the smaller
target aperture provided the correction factor needed to generate
a model airglow spectrum tailored to each target spectrum. Our template airglow
spectra were then corrected for slight wavelength offsets and scaled to
match the flux observed through the target aperture by linear shifting
and scaling to the Ly$\beta$\ line in the object spectrum.
The mean airglow-subtracted, exposure-weighted spectrum is shown in Figure 1.

\begin{figure*}[t]
\plotfiddle{"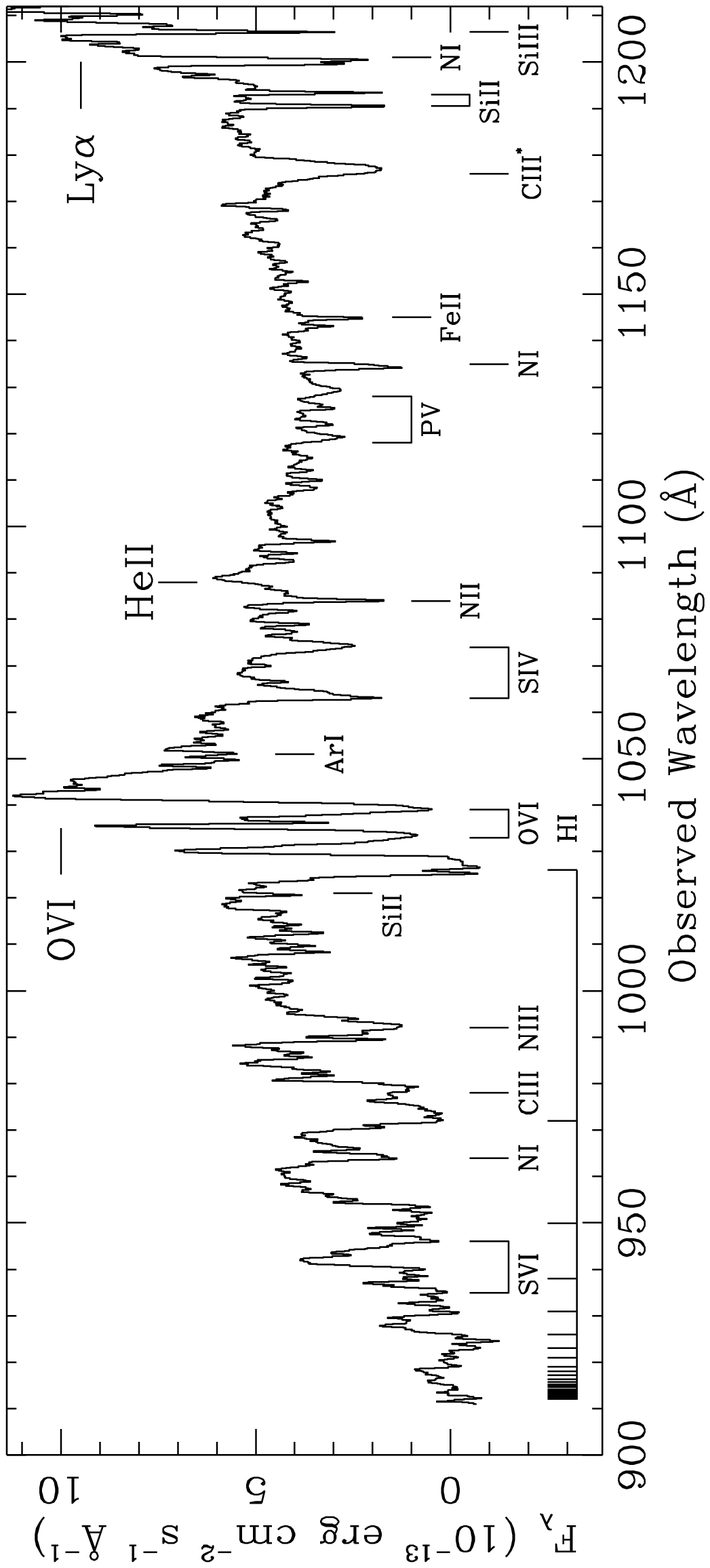"}{2.7in}
{-90}{69}{69}{-265}{398}
\vbox to 12pt{\vfill}
\parbox{7.25in}{
\vbox to 12pt{\vfill}
\small\baselineskip 9pt
\footnotesize
\indent
{\sc Fig.}~1.---
Observed mean airglow-subtracted spectrum of NGC~4151 obtained with
the Berkeley spectrometer on the ORFEUS-SPAS~II mission in
November/December 1996.
}
\end{figure*}

The mean spectrum shows that NGC~4151 had an intensity intermediate 
between the observations made during the Astro-1 and Astro-2 missions.
We use the mean spectrum to determine the basic parameters of both
continuum and line components (see Tables 1 and 2).
These and subsequent fits were performed using $\chi^2$ minimization in
the spectral fitting program {\tt specfit} (\cite{Kriss94}).
We model the continuum as a power law in $f_\lambda$.
We correct for foreground extinction using a
Clayton, Cardelli, \& Mathis (1989)\markcite{CCM89}
curve assuming $E(B-V) = 0.04$ (\cite{Kriss95}) and $R_V = 3.1$.

The best-fit, extinction-corrected continuum is
$f_\lambda = 8.20 \times 10^{-13} (\lambda/1000)^{-1.37}$.
For the Ly$\alpha$, \ion{He}{2} $\lambda 1085$ and \ion{O}{6} $\lambda 1036$
emission lines we assume the shape
found by Kriss et al. (1992)\markcite{Kriss92}.
Absorption features were modeled using components
that are either Gaussians in equivalent width or optical depth.
The features identified in the mean spectrum and their parameters are listed
in Table 1.
We find no evidence for a change in power-law index among the individual
observations.
To fit the variations in the individual emission and absorption lines,
we fix the wavelength and line width of each component and fit the intensity or
equivalent width in every individual exposure.

\section{ABSORPTION FEATURES AND THEIR VARIABILITY}

\subsection{Neutral Hydrogen Absorption}

Measuring the intrinsic neutral hydrogen column in NGC~4151 requires good S/N
in the high-order Lyman lines since the low-order lines are heavily saturated.
A S/N of $\sim 5$ per bin at the Lyman limit required further binning of the
the mean spectrum to 0.6 \AA\ pixel$^{-1}$.
This is still far better than the $\sim$3 \AA\ resolution of HUT.
Unfortunately, the high background

\begin{center}
\small
{\sc Table 1\\
Absorption lines in the mean spectrum of NGC 4151}
\vskip 4pt
\begin{tabular}{lcccc}
\hline
\hline
ID & EW    &  {\it v}$_{\sun}^{\rm a}$ & FWHM                & Origin$\rm ^{b}$\\
   & (\AA) & ($\rm km\,s^{-1}$) & ($\rm km\,s^{-1}$)  &             \\
\hline
SVI 933.38	& $3.96 \pm 1.14$ & 533 & $567 \pm 131$ & N \\
SVI 944.52 	& $3.60 \pm 1.00$ & 533 & $567 \pm 131$ & N \\
NI 963.99 	& $0.55 \pm 0.26$ & 533 & $567 \pm 131$ & N \\
CIII 977.02	& $3.69 \pm 1.32$ & 508 & $567 \pm 110$  & N \\
NIII 990.79	& $2.54 \pm 0.09$ & 499 & $1000 \pm 200$   & N \\
SiII 1020.70 	& $0.26 \pm 0.03$ & 10 & $142 \pm 23$ & G \\
OVI 1031.93	& $3.54 \pm 0.28$ & 455 & \ldots & N \\
OVI 1037.62	& $3.81 \pm 0.47$ & 455 & \ldots & N \\
ArI 1048.22	& $0.31 \pm 0.10$ & 838 & $277 \pm 120$ & N \\
SIV 1062.67 	& $1.49 \pm 0.09$ & 379 & $768 \pm 50$   & N \\
SIV 1073.28	& $1.28 \pm 0.09$ & 379 & $768 \pm 50$   & N \\
NII 1083.99	& $0.46 \pm 0.04$ &  46 & $146 \pm 15$   & N \\
PV 1117.98	& $0.60 \pm 0.21$ & 377 & $468 \pm 140$  & N \\ 
PV 1128.01	& $0.64 \pm 0.34$ & 377 & $468 \pm 140$  & N \\ 
FeII 1144.94	& $0.40 \pm 0.06$ &  18 & $158 \pm 28$   & G \\
CIII$^*$\ 1175.70 	& $2.32 \pm 0.12$ & 341 & $846 \pm 53$ & N \\
SiII 1190.42    & $0.58 \pm 0.03$ &  60 & $154 \pm 10$  & G \\
SiII 1193.29    & $0.58 \pm 0.03$ &  60 & $154 \pm 10$  & G \\
NI 1199.9	& $1.31 \pm 0.08$ & 103 & $393 \pm 29$  & G \\
SiIII 1206.50   & $0.52 \pm 0.04$ &  33 & $140 \pm 12$  & G \\
\hline
\end{tabular}
\parbox{3.5in}{
\footnotesize
$\rm ^{a}$The Galactic Si~II $\lambda$1190 and $\lambda$1193 lines
(with {\it v}$_{\sun} = 60~\rm km~s^{-1}$)
were used to align the wavelength scale.
Velocities are accurate to $\sim40~\rm km~s^{-1}$ (1$\sigma$).

$\rm ^{b}$N: lines intrinsic to NGC~4151; G: Galactic absorption lines.
}
\end{center}
\vskip 9pt

\noindent
at short wavelengths prevents a good measure
of the hydrogen column in the individual observations.
No significant variations
in the strong Ly$\alpha$ and Ly$\beta$ absorption
lines are seen, presumably because of their high optical depth.
The higher spectral resolution of the Berkeley spectrometer compared to HUT,
however, enables us to deblend the hydrogen absorption in the mean
spectrum and separate the complex assortment of Galactic and
intrinsic components that are present.

The outflowing \ion{H}{1} absorption in NGC~4151 brackets strong absorption by
neutral hydrogen in our own Galaxy.
Galactic \ion{H}{1} 21-cm emission observed toward NGC~4151 shows
multiple components with a total column density of
$\sim 2 \times 10^{20}~\rm cm^{-2}$
at a mean heliocentric velocity of
$60~\rm km~s^{-1}$ (\cite{Murphy96}; \cite{Murphy97}).
This complex structure is reflected in our ORFEUS-SPAS~II data by the large line
widths observed for Galactic absorption features.

To model the Galactic \ion{H}{1} absorption, we first fit the absorbing column,
location and line width of the four main components in Murphy's 21-cm profile.
We then generated an \ion{H}{1} absorption line template
smoothed to the resolution of our spectrum.
We find that although the total
absorbing column is $2 \times 10^{20}~\rm cm^{-2}$\ and the separation of the
components is less than our instrumental resolution, the resulting absorption
line profile is optically thick but not black, consistent with our observations
of the high-order Lyman lines.

\begin{table*}[b]
\begin{center}
\small
\parbox{7.25in}{
\begin{center}
\small
{\sc Table 2\\
Hi Absorption in NGC 4151}
\end{center}
}
\begin{tabular}{lcccc}
\hline
\hline
Component & {\it v}$_{\sun}$\ ($\rm km\,s^{-1}$) & log N(HI) ($\rm cm^{-2}$) & b ($\rm km\,s^{-1}$) & Covering fraction \\
\hline
NGC~4151 `A' & $-540$ & $15.5 \pm 0.1$   &  $45 \pm 0$ & $0.95 \pm 0.24$ \\
Galactic     & \phantom{$-0$}60     &  $20.3 \pm 0\phantom{.0}$     &  $75 \pm 0$ & $1.00 \pm 0\phantom{.00}$  \\
NGC~4151 `C+D+E' & \phantom{$-$}451  & $18.70 \pm 1.5$ & $250 \pm 50$ & $0.84 \pm 0.06$ \\
NGC~4151 `F'     & \phantom{$-$}982 & $19.61 \pm 0.5$ &  $45 \pm 0$ & $0.73 \pm 0.80$ \\
\hline
\end{tabular}
\end{center}
\end{table*}

We similarly determined
the location and line width of two narrow low-ionization components seen
in GHRS spectra (components A and F) which are identified with outflowing
and halo material in NGC~4151, respectively (\cite{Weymann97}).
We fit these components in the mean spectrum
and included a broad component to represent higher ionization material
(an unresolved blend of components C, D and E).
The several components and their best fit parameters are summarized in Table 2.
The reduced $\chi^2$\ of the fit is quite large (187.6),
but this is largely due to an underestimate of the true errors in the
binned data.  Near \ion{O}{6} in the re-binned spectrum the formal S/N
based on photon statistics is $\sim$85,
but the flat field is only accurate to $\sim 5$\%.

Our best fit to the broad absorption gives $\log (\rm N_{HI}) = 19.7$\ and 
$b = 200~\rm km~s^{-1}$ covering $\sim 84$\% of the source.
This broad component is consistent with that seen in the HUT Astro-1 and
Astro-2 spectra (\cite{Kriss92}; \cite{Kriss95}).
It has a width and partial covering similar to that
seen during Astro-2, and a total column intermediate between the Astro-1
and Astro-2 values, consistent with the intermediate value of the continuum
flux seen in these new observations.
Component `F', at the systemic velocity of NGC~4151 and probably associated
with its ISM or halo, has a column density an order of magnitude lower than
the optically thick component suggested by
Kriss et al. (1995)\markcite{Kriss95},
but it is still optically thick.
Given the lack of broad, black \ion{Mg}{2} absorption at this velocity in the
GHRS spectrum (\cite{Weymann97}),
it is likely that the actual column density is at the lower
end of the range permitted by our errors.

\subsection{Other Absorption Features}

Correcting for the $90~\rm km~s^{-1}$ resolution of the Berkeley spectrometer,
the unresolved Galactic absorption features have FWHM $\sim 100~\rm km\,s^{-1}$
(see Table 1).
This is roughly what is expected from the 21-cm
\ion{H}{1} profile towards NGC~4151 (\cite{Murphy97}), and
is consistent with Galactic lines in the GHRS spectra.
Our identified Galactic absorption lines are similar to those reported for
the 3C~273 sight line by Hurwitz et al.\ (1997b).\markcite{Hurwitz97b}

Accurate measurements of \ion{O}{6} absorption intrinsic to NGC~4151 proved
difficult due to the overlapping absorption and emission features.
What is apparent, however, is that the dominant absorption in the \ion{O}{6}
doublet is similar to the absorption features seen in the \ion{C}{4} line
(\cite{Weymann97}; \cite{Hutchings97}).
It is optically thick, but the broad width ($\sim1100~\rm km~s^{-1}$) and
the non-black line centers imply partial covering.
We see no significant variations in the \ion{O}{6} absorption equivalent width.
More detailed analysis of the \ion{O}{6} troughs will be reported in a
future publication.

As described by Bromage et al. (1985)\markcite{Bromage85} and
Kriss et al. (1992)\markcite{Kriss92}, \ion{C}{3}$^*$\ $\lambda$1176 can be used
together with \ion{C}{3} $\lambda$977 to measure the density of the
absorbing gas.
In the Astro-1 and Astro-2 HUT data, both lines were assumed to arise in
the same absorbing region.  The higher resolution ORFEUS-SPAS~II data
shows that the situation is not so simple--- the mean velocity of the
\ion{C}{3}$^*$\ $\lambda$1176 absorption is significantly higher than
that of \ion{C}{3} $\lambda$977.
\ion{C}{3} $\lambda$977 is less blueshifted, and it has a velocity
comparable to that of component D in the \ion{C}{4} profile of
Weymann et al. (1997)\markcite{Weymann97}.
\ion{C}{3}$^*$\ $\lambda$1176 is closer in velocity to component C,
but the 850 $\rm km~s^{-1}$ width of the feature gives considerable
overlap to all broad \ion{C}{4} components C, D, and E.
If the gas outflowing from NGC~4151 is accelerating,
this places the highest density gas in the outflow farthest from the source.

In contrast to the results for Ly$\beta$\ and \ion{O}{6} absorption,
we detect strong variations in the \ion{C}{3}$^*$\ $\lambda$1176
and \ion{S}{4} $\lambda$1073 absorption lines.
(\ion{S}{4} $\lambda$1062 is blended with Galactic absorption.)
The constancy of the Galactic features throughout the mission confirms
the reality of these variations.
As shown in Figure \ref{n4151f2.eps}, both \ion{C}{3}$^*$\ and 
\ion{S}{4} anti-correlate with the observed continuum variations.
To within the 1--2 day sampling of our measurements, the continuum and
line variations are instantaneous with no significant lag.

Following the formalism of Krolik \& Kriss (1997)\markcite{KK97},
we can use the ionization and recombination timescales
to set limits on the location of the absorbing material.
For an observed ionization timescale $t_{ion}$, an ionizing flux density
at the earth $f_{ion}$,
a mean photoionization cross section $\langle \sigma_{ion} \rangle$,
an ionization energy $h \nu_T$, and a distance to the source $D$ (15 Mpc for
NGC~4151), they find that the material must lie at a radius
\begin{equation}
r \leq \left({f_{ion} \langle \sigma_{ion}\rangle t_{ion} \over
       h\nu_T}\right)^{1/2} D.
\end{equation}
The decrease in the \ion{C}{3}$^*~\lambda1176$ equivalent width
during the increase in the continuum flux observed from day 334 to 335
implies that ionization is dominating in this interval.
We thus have an upper limit of 1 day on $t_{ion}$.
Using a photoionization cross section of
$\langle \sigma_{ion} \rangle = 1.6 \times 10^{-18}~\rm cm^2$
(\cite{Osterbrock89})
at the ionization threshold for \ion{C}{3}$^*$\ (41.4 eV, or 300 \AA),
and extrapolating our best-fit, extinction corrected continuum to 300 \AA,
we find $r < 25$ pc.
This is comparable to the limit found using the variation in the
\ion{H}{1} column seen during the HUT Astro-2 observations of NGC~4151
(\cite{KK97}; \cite{Kriss98}).

\section{EMISSION FEATURES}
 
The dominant emission features in our data are \ion{O}{6} $\lambda$1034,
\ion{He}{2} $\lambda$1085, and the blue wing of Ly$\alpha$.
The partial Ly$\alpha$\ line and the
strength of the Ly$\alpha$\ airglow preclude any meaningful
determination of the intrinsic line strength.
Variations in the \ion{O}{6} and \ion{He}{2} fluxes compared to the continuum
are shown in Figure \ref{n4151f2.eps}.
The fractional variation in the \ion{He}{2} flux ($F_{var} = 0.23$) is
comparable to that in the continuum flux ($F_{var}(1000) = 0.20$).
In contrast, the variation seen in the \ion{O}{6} emission is much less:
$F_{var} = 0.12$.
Our data set is not large enough to compute meaningful cross correlation
functions, but qualitatively one can see that there is significant response
in the emission lines on timescales of 2 days or less.
The detailed response, however, is not the typical smeared and delayed
variation seen in prior monitoring campaigns.
While the continuum is rising in the first three observations,
\ion{O}{6} and \ion{He}{2} are constant or falling in flux.
In contrast, both lines track well the dip and subsequent rise in continuum flux
at days 334 to 336.

\acknowledgements
We thank the Berkeley spectrometer science team and all
those whose participation helped make the ORFEUS-SPAS~II mission
such a success. This work was supported by NASA LTSA grant number
NAG 5-3255 to the Johns Hopkins University.

\vbox to 14pt{\vfill}
\vbox to 5.1in {
\plotfiddle{"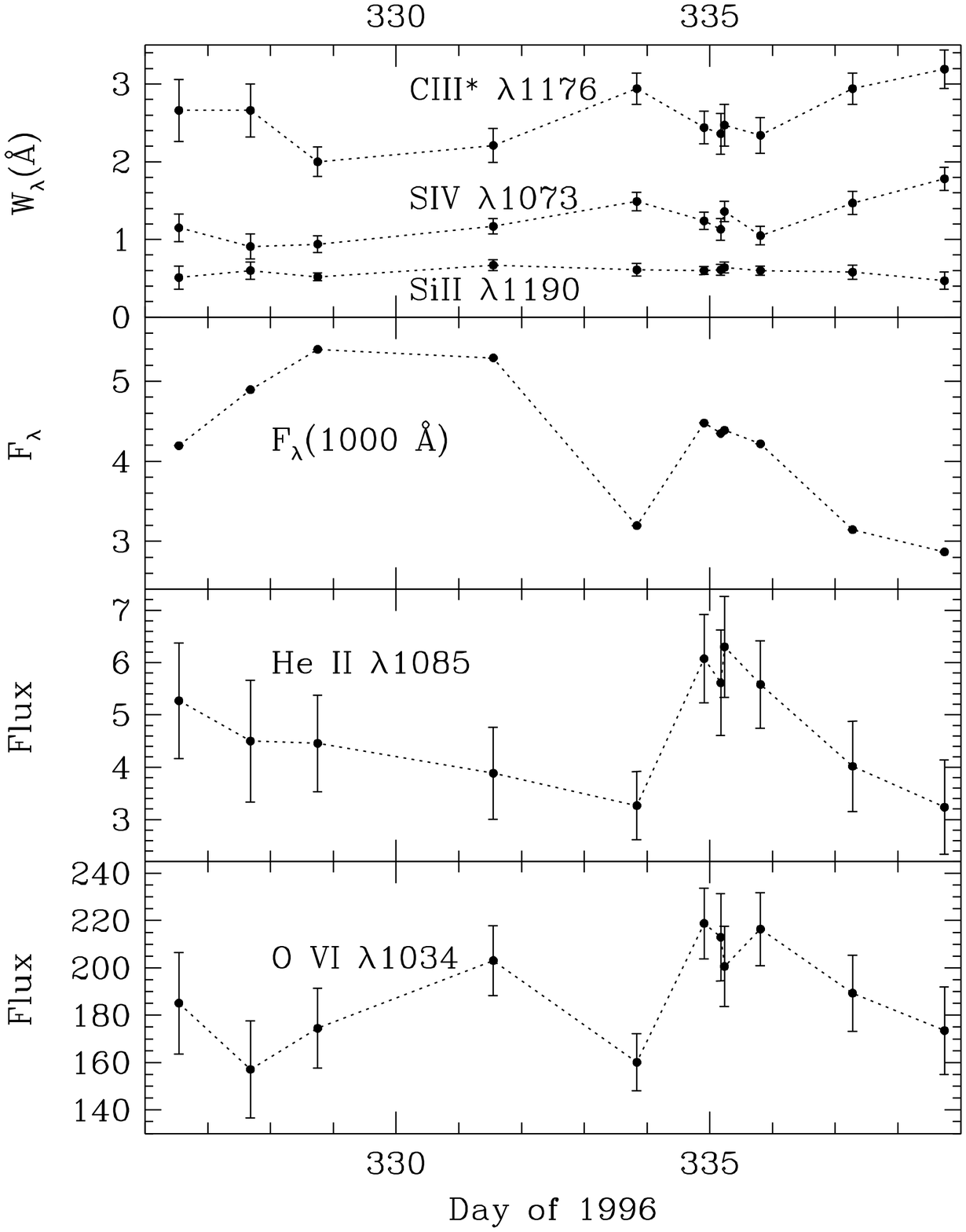"}{4.3in}
{0}{45}{45}{-137}{-20}
\parbox{3.5in}{
\small\baselineskip 9pt
\footnotesize
\indent
{\sc Fig.}~2.---
Absorption line, continuum, and emission line variations with time
for the eleven observations of NGC~4151 during the ORFEUS-SPAS~II mission.
Note that the Galactic Si~II $\lambda 1190$ equivalent width in the top panel
shows no evidence for variation.
The continuum flux at 1000 \AA\ is in units
of $10^{-13}~\rm erg~cm^{-2}~s^{-1}~\AA^{-1}$.
The He~II $\lambda$1085 and O~VI $\lambda$1034 emission line fluxes are in
units of $10^{-13}~\rm erg~cm^{-2}~s^{-1}$.
\label{n4151f2.eps}
}
}

\end{document}